# Superatom Representation of High-$T_C$ Superconductivity

Itai Panas

Environmental Inorganic Chemistry

Division of Energy & Materials

Department of Chemistry and Biotechnology

Chalmers University of Technology

S-412 96 Gothenburg, Sweden

**Abstract**

A "super-atom" conceptual interface between chemistry and physics is proposed in order to assist in the search for higher $T_C$ superconductors. High-$T_C$ superconductivity HTSC is articulated as the entanglement of two disjoint electronic manifolds in the vicinity of a common Fermi energy. The resulting HTSC ground state couples near-degenerate protected local "super-atom" states to virtual magnons in an antiferromagnetic AFM embedding. The composite Cooper pairs emerge as the interaction particles for virtual magnons mediated "self-coherent entanglement" of super-atom states. A Hückel type resonating valence bond RVB formalism is employed in order to illustrate the real-space Cooper pairs as well as their delocalization and Bose Einstein condensation BEC on a ring of super-atoms. The chemical potential $\mu_{BEC}$ for Cooper pairs joining the condensate is formulated in terms of the super-exchange interaction, and consequently the $T_C$ in terms of the Neél temperature. A rationale for the robustness of the HTSC ground state is proposed: achieving local maximum "electron correlation entropy" at the expense of non-local phase rigidity.

## 1. Introduction

A quarter of a century has passed since the discovery of high critical temperature superconductivity HTSC in the cuprates, see e.g. [1-6]. From the perspective of the theoretical chemist, it becomes important to provide a chemically simple yet sufficiently complex conceptual frame work [7-9] to invite chemical manipulations as well as quantum chemical model calculations [10-13] offering guidance in the search for materials with higher critical temperatures. A central message is the form, which a viable mechanistic understanding of HTSC must take in order to be useful from a solid state chemistry perspective, here in terms of super-atoms exhibiting protected accidental degeneracies and embedded in an antiferromagnetic surrounding. This conceptual understanding [7,8] has been explored during the last decade. Thus the predicted checkerboard structure [7,12,14-17], presence and essential roles of short-range AFM fluctuations [18,19] and in particular the magnetic spin-flip excitation [7,12,20-22], c-axis dynamics of the ions at the A-sites [7,9,10,23], d-wave order parameter symmetry [7,24] and a fine local energy scale in vortex cores [7-13,25] have all been confirmed by experiment as being essential. In particular, these observations are all consistent with an understanding, which has the HTSC develop from a pseudogapped state [26], and contradict scenarios based on HTSC emerging from a well developed Fermi surface [27,28].

With the recent discovery of superconductivity in the Fe-chalcogenide [29] and Fe-pnictide [30] materials, there is a growing impression in the solid state community that there exists a possible general underlying principle for achieving SC in the so-called strongly correlated systems, among which the cuprates and the iron based materials are but two representatives of the class. In this context a common such possible understanding was formulated in ref. [31].

Today it is acknowledged that the Bardeen Cooper Schrieffer BCS theory for superconductivity [32] is but the simplest possible microscopic formulation of the phenomenon, as it succeeds in demonstrating how an electron gas model system may undergo transition into a superconductor. However, taking the electron-gas Hamiltonian as 0:th order ansatz is in general not physical, and particularly not so for so-called strongly correlated systems, such as the cuprates and the new iron based superconductors. Arguably, it was the identification of the SC phenomenon with its BCS theory representation, i.e. transition into the superconducting state from an electron gas reference state, which rendered particular interest to the discovery of HTSC in the strongly correlated systems. Materials which display strong correlations were thought to offer all the reasons for not turning superconducting, such as spin-density waves, charge-density waves, giant magneto-resistance, colossal magneto resistance, orbital ordering, ferromagnetic metallicity, antiferromagnetism etc. These properties are all signatures of instabilities, which materialize by corresponding symmetry breakings. Yet, in the end it was in the strongly correlated systems where the highest critical temperatures for superconductivity were found, thus underlining the implicit paradigm shift required to come to terms with the HTSC phenomenon. In [7,31] it was proposed that what distinguishes the superconductors among the strongly correlated systems is that when all "normal" symmetry breakings have occurred, of which short-range AF order is essential, in the HTSC materials protected local electronic degeneracies remain to be lifted by the superconductivity.

In what follows, entering the SC state is achieved in two steps. Thus, the so-called pseudogap is proposed to reflect an instability of semi-local origin, which materializes as "pre-formed" real-space Cooper pairs, i.e. the protected accidental degeneracies in the manifold of super-atom states, in the hole-doped cuprates or among local d-states in the Fe-superconductors, are consumed by non-adiabatic coupling between short-range AFM and super-atom virtual excitations producing the pseudogap. Many of the superconducting properties are indeed associated with the virtual magnons in the antiferromagnet resulting from this coupling. Means for the system to lower its energy further is offered by the Bose Einstein condensation of said real-space Cooper pairs, the consequence of which is the non-local phase rigidity of HTSC.

## 2. Super-atom as composite conceptual building block for superconductivity

While this super-atom approach is novel in the context of HTSC, the resulting overall understanding shares essential common flavors with the Resonating Valence Bond RVB model of Anderson [33-35]. That theory takes spin-charge separation in a doped Mott-Hubbard insulator as point of departure, and because the charge excitations do not carry spin, they may undergo Bose-Einstein condensation. A key feature of that theory is almost-free holes amplitudes displaying the symmetry of bound electron pairs. Similarly, our quantum chemistry based conceptual model achieves said “separation” by the charge-carriers physically separating from the AFM band, thus forming hole clusters, i.e. “super-atoms”. The HTSC emerges from a resonance requirement for virtual excitations in the antiferromagnet with virtual excitations in the super-atoms. Indeed, this understanding offers an alternative frame work for the Anderson RVB spinon-holon phenomenology. In the latter, holes-pairs amplitudes are said to become complementary to the "bound electrons" such that the former may undergo Bose Einstein condensation. Below, this is realized explicitly in the super-atoms representation.

The chemical perspective assumed in the present study is based on local electronic motifs, i.e. "super-atoms" and that superconductivity emerges upon achieving electronic phase coherence among such objects. Thus it bears significant resemblance to heavy fermion superconductivity [36]. In [7-13] charge carriers clustering was discussed and demonstrated to occur as a response to disorder in the position of the ions at the A-site. Let these hole clusters (see Figure 1) constitute said "super-atoms" and interpret the sharing of a pair-state among precisely two such "super-atoms" to reflect the formation of a real-space Cooper pair. Such an ansatz builds on locality constraints of non-adiabatic coupling in two electronic sub-systems comprising virtual excitations in super-atoms manifolds mediated by virtual magnons in the local AFM background. Note e.g. in Figure 1b how the AFM band characteristics, seen best in the spin density on the rim of the 4 x 4 unit cells superstructure, comes out different from the "super-atom" spin density in the center. This observation offers an illustration of our dual formulation of superconductivity, which explains our deviation from the RVB understanding in its simplest form. Hence, it deviates from the charge-spin separation in a single band, as charge carriers are allowed to leave the AFM band altogether (see Figure 1 again) to build local super-atoms in terms of states that belong to a low-dispersive second band, see [7,31,37]. The set of non-bonding in-plane oxygen orbitals of $\pi$ symmetry with respect to the local Cu-O-Cu axis, are found to act as sinks for holes.

**Figure 1**

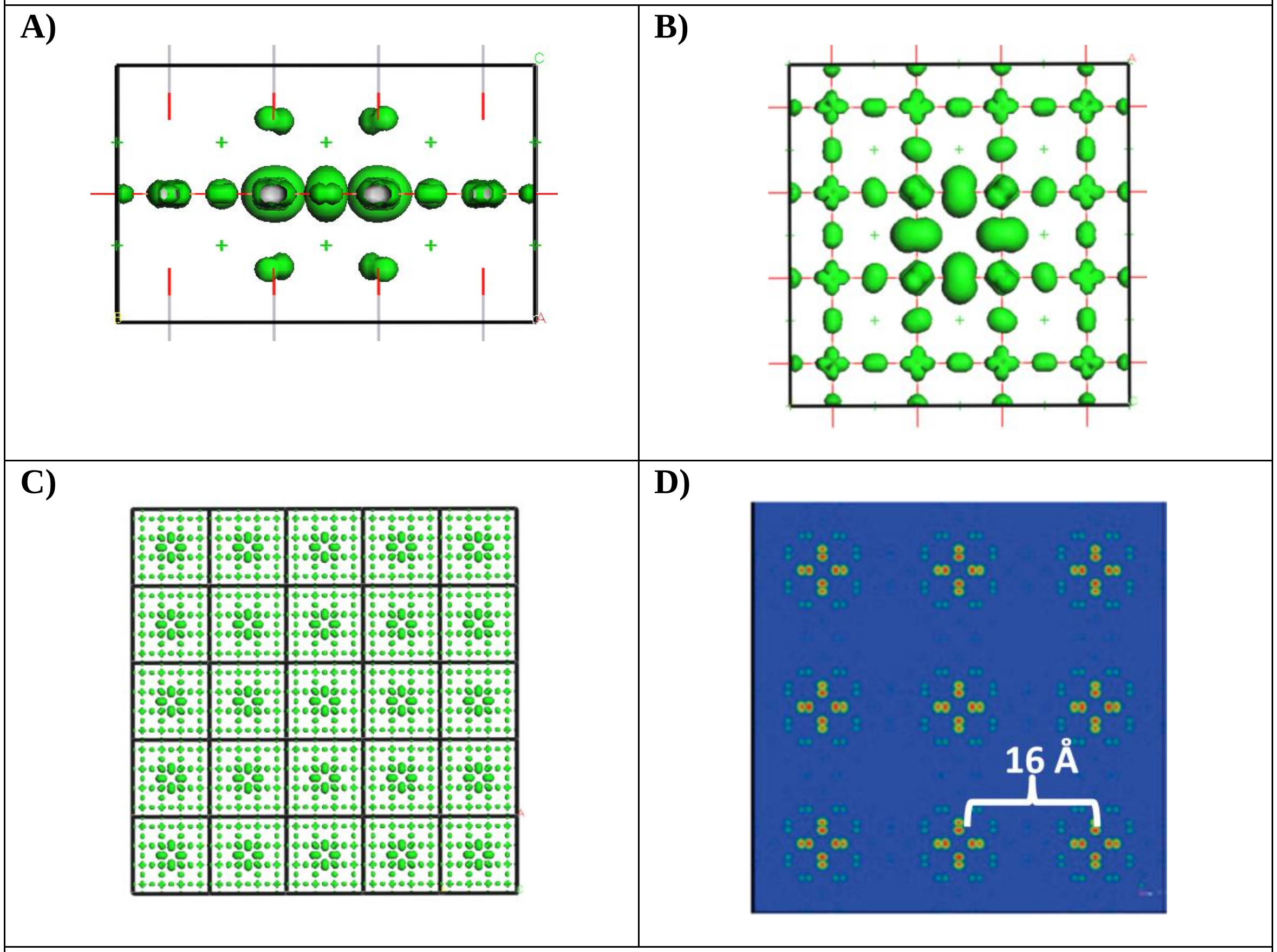

**Figure 1.** (A-C) DFT Spin densities in $HgBa_{1.87}CuO_4$ for 25% holes doping produced by the introduction of 2 Ba vacancies in $4a_0 \times 4a_0$ unit cells superstructure. (A) and (B) are side views and top views respectively. (C) emphasizes the checkerboard structure providing possible connection to STM experiments [14-16]. (D) depicts complementary sums-over-states-densities for periodic hole clusters due to central $\delta$-O in the Hg plane, producing a $3 \times 3$ super-cells for Hg-1201, from [36].

In what fllows, an RVB interpretation of the HTSC, complementary to that employed by Anderson is articulated, although real-space Cooper pairs are indeed said to form and HTSC accessed by subsequent Bose Einstein condensation. Let the two wave functions $\phi_1$ and $\phi_2$ represent local super-atoms at site 1 and site 2, where $5 \times 5$ $(3 \times 3)$ such sites are shown in Figure 1C (1E). Let the ground state of the super-atom be a node-less pair-state, which implies that the first excited state is pair-broken and orthogonal to the pair. We postulate that formation of a Cooper pair is equivalent to writing the local ground state wave function at any site ***i*** as a superposition of super-atom pair (P) and pair-broken (PB) states

$$\phi_i = u_i \cdot |PB_i> + v_i \cdot |P_i> \qquad (1)$$

It implies that a real-space Cooper pair is a shared pair-state among two superatoms. This local wave function ansatz has been exemplified and discussed in some depth [7,31]. It implies the lifting spin and symmetry constraints, such that super-atom pair- and pair-broken states are allowed to mix. This mixing can be understood to reflect an "entropy" among electron configurations as inspired by Shannon's information theory [38], i.e. "delocalization" in the space spanned by all accessible near-degenerate local electronic states. This "disorder" is made possible by the coexistence of a second electronic subsystem acting as spin and symmetry buffer, displaying complementary local virtual magnons in local antiferromagnetic an embedding, matching the virtual super-atom excitations. The corresponding "local internal space" is discussed in [7]. Thus, neither super-atom nor local AFM channels preserve local spin and space symmetry. Instead local spin and space symmetry is preserved locally for the local $|AFM> \otimes |Superatom>$ compound. Access to the local maximum-entropy super-atom states requires phase coherence among the super-atoms such that the ground state reflects non-local entanglement of these locally incommensurate states by the complementary mixing of incommensurate magnetic states comprising virtual magnons injected into the AFM. Cooperative phase rigidity in the combined manifolds of virtual super-atoms excitations and virtual magnons is required in order to maintain the local entanglements at sites A and B. For a diagrammatic representation of the resulting entanglement, see Figure 2. This inter-system phase locking between said two virtual

**Figure 2**

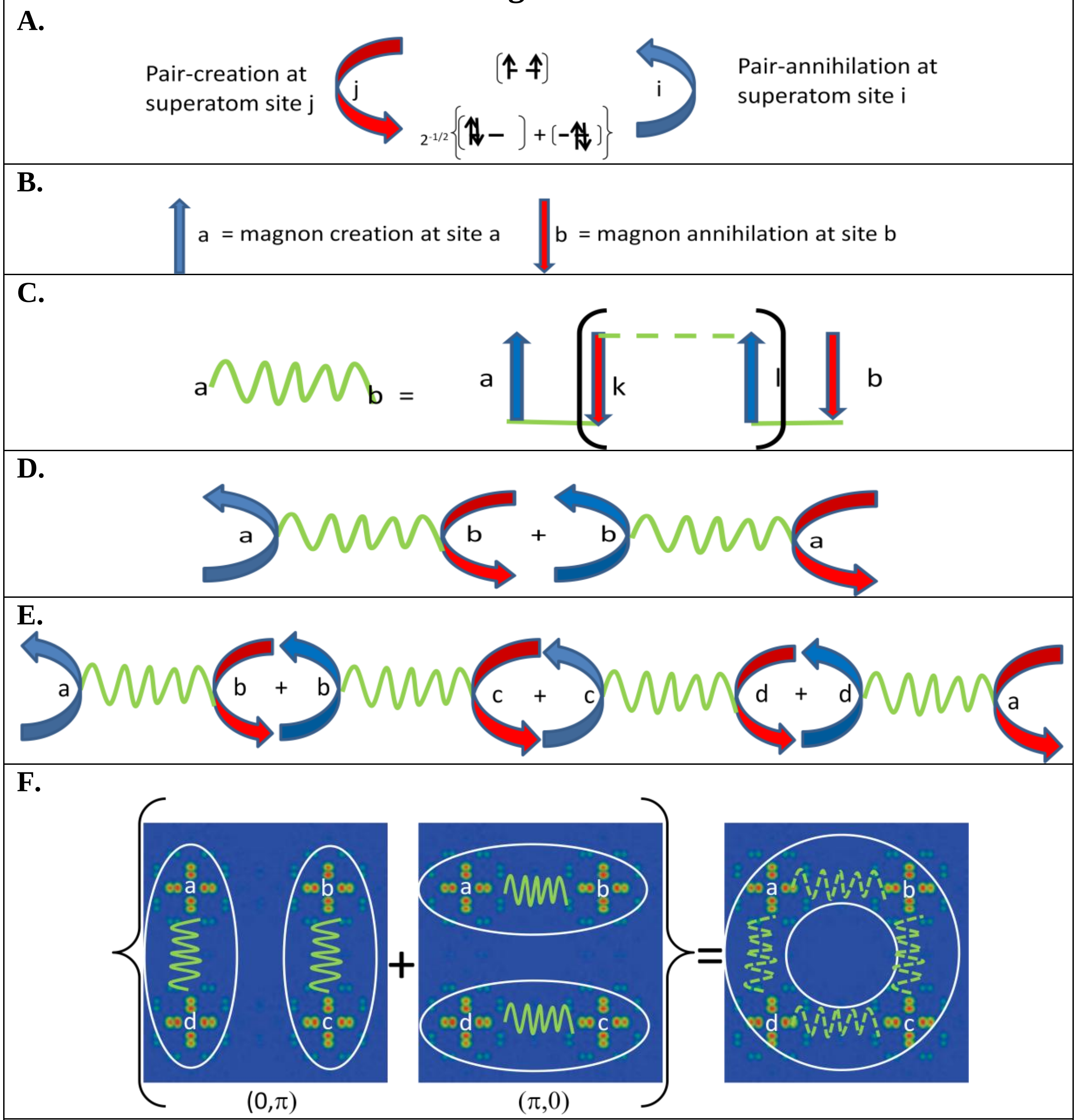

**Figure 2. A**) Red curved arrow annihilates a pair-broken super-atom state and creates a pair-state, while blue arrow represents the inverse process. **B**) Blue straight arrow creates a magnon, while red arrow annihilates a magnon. **C**) Green wiggly line transfers magnon from site **a** to site **b** via intermediates sites **k,..,l.** (**D**) Entanglement at site **a** is made possible by entanglement at site **b** as mediated by shuttling a magnon between the two sites. **E)** As in D) but emphasizing the near-sightedness of non-local phase locking required for BEC. **F)** as E), but emphasizes how superposition of two local conjugated Cooper pairs, shared among four super-atoms a-d, display gaps in the $(0,\pi)$ and $(\pi,0)$ directions allows for effective delocalization on a ring thus achieving the BEC.

BEC:s may be understood from an extended electronic liquid crystal stripes CDW perspective, see [39] for possible smectic electronic order. Arrangements of so-called "conjugated" Cooper pairs (see below) can be understood to produce such liquid crystal order, and while this order breaks fourfold rotational symmetry, tunneling between two stripes directions at $90^{o}$ angle. A gap opens owing to this tunneling, and a complementary tunneling gap opens in the AFM manifold of states. see Fig. 2F again for a local perspective on how the composite system recovers the underlying four-fold symmetry as a superposition of two symmetry broken subsystems. The resulting cooperative ground state is associated with an effective order parameter of the form

$$O(\chi, \varphi) = |\chi||\varphi| \cdot e^{i\theta} \qquad (2)$$

where $\chi$ and $\varphi$ are the complex BEC order parameters of the virtual magnons and super-atom excitations, respectively, while $\theta$ is the relative phase between the two cooperating condensates. Here, the composite Cooper pairs comprise the interaction particles for virtual magnons mediated "self-coherent entanglement" of virtual super-atom states.

It is noted carefully that the origin of the angular dependence of the gap functions of the superconducting and pseudogapped states may be traced to the resolution of the delocalized ring arrangement (see yet again Fig. 2F), i.e. in the form of superpositions of resonances in the $(0, \pi)$ and $(\pi, 0)$ directions in case of superconducting state, and as a statistical average of the two directions in the pseudogapped state.

### 3. Making contact between super-atoms and superconductivity

Taking our departure from the composite ground state as reflected in the order parameter, see Eq. 2, let us now focus entirely on the super-atom perspective. Thus, we exploit the postulated protected accidental degeneracy in order to write the local spin and symmetry violating super-atom wave function as a superposition of chemical pair- and a pair-broken states for each individual super-atom, see Eq.1, and note that each of $|PB_i>$ and $|P_i>$ comprises exact correlated eigenstates to the same local many-electron hamiltonian $H_i^0$, such that in this sense the two states are orthogonal. For *N* such non-interacting super-atoms NIS, the macroscopic wave function becomes

$$\Psi_0^{NIS} = \prod_{i=1}^{N} \phi_i \qquad (3)$$

Repeatedly, this becomes allowed by the particular non-adiabatic coupling to the AFM manifold of near-degenerate electronic states. Hence, the necessary requirement for a super-atom representation of HTSC is that there is an additional field, $H_i^{'}$, which allows local spin and space symmetry to be violated on each super-atom. Assume now that this background field carries a "stiffness" as represented by local antiferromagnetic order. In this case, the local field $H_i^{'}$ is replaced by a coupling hamiltonian $H_{ij}^{'}$ which conditions the correlated state (1) in such a way that *two super-atoms share a pair-state*, i.e. $u_j^2 = v_i^2$. Let pair-wise interactions between super-atoms delocalize the pair-state according to a coupling matrix element of the form

$$\langle H_{12}\rangle = \langle \phi_2 | C^{+}_{2,P_0} C_{2,PB_1} | \phi_2 \rangle \times \langle \Psi_M | M^{+}_{2,P_0} M_{2,PB_{-1}} | W | M^{+}_{1,PB_{-1}} M_{1,P_0} | \Psi_M \rangle \times \langle \phi_1 | C^{+}_{1,PB_1} C_{1,P_0} | \phi_1 \rangle \quad (4)$$

See again Eq. 1 for definition of $\phi_i$. $C_{i,\Sigma_S}$ $(M_{i,\Sigma_S})$ annihilates a local state of symmetry $\Sigma$, i.e. *P* or *PB* in superatom (AF/magnon in the AFM ) with projected spin $S_Z$=0,$\pm$1 in the super-atom (AFM). $C^{+}{}_{i,\Sigma_S}$ $(M^{+}{}_{i,\Sigma_S})$ creates corresponding states. Thus, Eq. 4 describes how a local transition from the pair-state into the pair-broken state in one super-atom, is accompanied by the complementary transition from a pair state into a pair-broken state in the magnetic channel in the vicinity of that site, while preserving the local compound electronic spin and space symmetry. This is in turn accompanied by the precise complementary process at a second site, thus fulfilling the non-local entanglement condition (cf. Fig. 2A-D). By introducing Eq. 1 into Eq.4 we obtain

$$\beta \equiv H_{12} = v_1 \cdot u_1 \cdot u_2 \cdot v_2 \cdot J_{local} \qquad (5)$$

where $J_{local}$ is a measure of the AFM stiffness required in order to maintain the non-local entanglement. This is an attractive interaction which can be understood as a trade-off between the zero entropy non-local phase rigidity of the macroscopic wave function, and the maximum "electron correlation entropy" among internal super-atom states, in that both pair- and pair broken local internal superatom states contribute to the ground state. The fact that the fidelity of this entanglement is related to the magnitude of the super-exchange interaction lends crucial significance to incommensurate magnetic fluctuations [19] and magnetic spin-flip resonance in inelastic neutron scattering [20, 21], observed at $T \leq T_C$, consistent with recent μ-spin resonance measurements [22].

Now we first consider a superatom dimer. For simplicity assume

$$u_1 = u_2 = u$$

$$v_1 = v_2 = v$$

$$\alpha_i = \alpha$$

and solve the Schrödinger equation for the wave function ansatz

$$\Phi_{\pm} = c_1\phi_1 \pm c_2\phi_2 \qquad (6)$$

A Hückel type hamiltonian for interacting super-atoms, *c.f.* Eq. 5, results in the secular equation

$$\begin{vmatrix} \alpha & -E & \beta & \\ \beta & & \alpha & -E \end{vmatrix} = 0 \qquad (7)$$

the solutions of which offer the eigenvalues and eigenstates

$$E = \alpha \pm \beta \text{ and } c_1 = c_2 = \frac{1}{\sqrt{2}} \qquad (8)$$

i.e. Cooper pair formation is associated with an energy gain of -2$\beta$. We extend this simple result by applying it to super-atoms on a ring, taking the pair-sharing to reflect formation of a nearest neighbor resonating valence bond, see any of the terms in brackets in Fig. 2E & 2F, taken to reflects a tight binding wave function corresponding to local independent conjugated Cooper pairs ICCP:s on a ring. For $\left[\frac{N}{2}\right]$ such decoupled local RVB pairs, corresponding to

$$\hat{H}_{RVB}^{ICCP} = \sum_{\alpha=1}^{[\frac{N}{2}]} (\widehat{H}_{2\alpha-1}^{0} + \hat{H}_{2\alpha}^{0} + \hat{H}_{2\alpha-1,2\alpha}^{'}) \qquad (9)$$

the Peierls type charge density wave becomes

$$\Psi_{RVB}^{ICCP} = \prod_{\alpha=1}^{[\frac{N}{2}]} \frac{1}{\sqrt{2}} (\phi_{2\alpha-1} + \phi_{2\alpha}) \qquad (10)$$

Equivalent to Eqs. (6-8), for a system of non-interacting Cooper pairs the corresponds secular equation becomes

$$\begin{vmatrix} \alpha - E & \beta & 0 & 0 & . & . \\ \beta & \alpha - E & 0 & 0 & . & . \\ 0 & 0 & \alpha - E & \beta & & . \\ 0 & 0 & \beta & \alpha - E & . & . \\ . & . & . & . & . & . \\ . & . & . & . & . & . \end{vmatrix} = 0 \qquad (11)$$

such that the resulting local resonating valence bond RVB between two adjacent super-atoms is represented by

$$\frac{1}{\sqrt{2}}(\phi_{2\alpha-1} + \phi_{2\alpha}) \qquad (12a)$$

and the resulting ground state energy is simply

$$NE_0^{ICCP} = N(\alpha + \beta) \qquad (12b)$$

Now, the simplest way to achieve Bose-Einstein condensation of *a priori* local ICCP:s is to offer any super-atom on the ring *the same interaction with both its nearest-neighbors left and right*, thus overcoming the above Peierls-type CDW, see Eqs. (9-12) . For a diagrammatic representation of a 4-superatoms ring, see again Figs. 2E & 2F. The corresponding hamiltonian comprises

$$\hat{H}_{RVB}^{BEC-CCP} = \sum_{i=1}^{N} (\hat{H}_i^0 + \hat{H}'_{i-1,i}) \qquad (13)$$

where the index 0 for $i = 1$ in $H'_{01}$ refers to the N:th super-atom in the cyclic arrangement. Now, the corresponding Hückel-type secular equation takes the familiar form

$$\begin{vmatrix} \alpha - E & \beta & 0 & 0 & . & \beta \\ \beta & \alpha - E & \beta & 0 & . & 0 \\ 0 & \beta & \alpha - E & \beta & & 0 \\ . & . & . & . & . & . \\ 0 & 0 & 0 & \beta & \alpha - E & \beta \\ \beta & 0 & 0 & 0 & \beta & \alpha - E \end{vmatrix} = 0 \qquad (14)$$

from which the well-known eigenvalues and eigenstates are obtained

$$E_k = \alpha + 2\beta \cdot \cos\left(k\frac{\pi}{N}\right) \qquad k = 0, \pm 1, \pm 2 \ldots, \pm(N-1), N \qquad (15a)$$

$$\Phi_k = \frac{1}{\sqrt{N}} \sum_{j=1}^{N} \exp\left[i \cdot k \cdot \pi \frac{(j-1)}{N}\right] \cdot \phi_j \qquad (15b)$$

Because the real-space Cooper pairs obey Bose-Einstein statistics, the BEC ground state has them all occupying the same delocalized single-boson ground state

$$\Phi_0^{BEC-CCP} = \frac{1}{\sqrt{N}} \sum_{i=1}^{N} \phi_i \qquad (15c)$$

where again $\phi_i = u_i \cdot | PB_i > + v_i \cdot | P_i >$ is the super-atom wave function at site *i*. The total energy for a condensate composed of $\left[\frac{N}{2}\right]$ such Cooper pairs becomes (cf. Eq. 15a for *k=0*)

$$NE_0^{BEC-CCP} = N(\alpha + 2\beta) \qquad (15d)$$

Comparing the stability of the Bose Einstein condensate of conjugated Cooper pairs (15d) to the stability of independent local conjugated Cooper pairs residing on $\left[\frac{N}{2}\right]$ super-atom dimers (Eq.12b) we conclude that the BEC indeed offers additional resonance stabilization, amounting to $2\beta$ per Cooper pair ($\beta$ per super-atom). It is emphasized that BEC resonance stabilizations, achieved upon including additional Cooper pairs in the condensate, are independent of number of Cooper pairs already present in the condensate, and that this BEC stabilization is mediated by complementary phase coherent magnons in the AFM embedding.

Given that the Cooper pairs belonging to the BEC are indistinguishable and that all pairs access the same delocalized single-Cooper-pair wave function, the wave function for the BEC is arrived at

$$\Psi_{\mathrm{RVB}}^{\mathrm{BEC-CCP}} = \prod_{\mathrm{n}=1}^{\left[\frac{\mathrm{N}}{2}\right]} \left( \frac{1}{\sqrt{N}} \sum_{\substack{i=1 \\ i \in BEC}}^{N} \phi_i\left(\vec{r}_n\right) \right) (16)$$

In case of finite temperatures, the various wave functions representing $M$ disconnected Cooper pairs coexisting with $\left[\frac{N}{2}\right] - M$ Cooper pairs in the condensate, take the form

$$\Psi_{\mathrm{RVB-mixed}}^{\mathrm{BEC-CCP}} = \prod_{\substack{\alpha=1 \\ \alpha \notin \mathrm{BEC}}}^{M} \frac{1}{\sqrt{2}} \left[\phi_{2\alpha-1}(\vec{r}_\alpha) + \phi_{2\alpha}(\vec{r}_\alpha)\right]$$

$$\times \prod_{\mathrm{n}=1}^{\left[\frac{\mathrm{N}}{2}\right]-\mathrm{M}} \left( \frac{1}{\sqrt{N-2M}} \sum_{\substack{i=1 \\ i \in BEC}}^{N-2M} \phi_i\left(\vec{r}_n\right) \right) \quad (17)$$

The mixed wave function (17) reflects the mixed hamiltonian

$$\hat{H}_{RVB-mixed}^{BEC-CCP}(T) = \sum_{\substack{\alpha=1 \\ \alpha \notin BEC}}^{M} \hat{H}_{2\alpha-1}^{0} + \hat{H}_{2\alpha}^{0} + \hat{H}_{2\alpha-1,2\alpha}^{'}) + \sum_{\substack{i=1 \\ i \in BEC}}^{N-2M} (\hat{H}_i^0 + \hat{H}_{i-1,i}^{'}) \qquad (18)$$

The expectation value for the number of disconnected Cooper pairs $M$ in such a thermally induced mixed state is determined by the Bose Einstein statistics, i.e.

$$M = [\frac{N}{2}] \times \frac{\sum_i \left[ \frac{1}{e^{\frac{\Delta E_i}{k_B T}} - 1} \right]}{\sum_i \left[ \frac{1}{e^{\frac{\Delta E_i}{k_B T_C}} - 1} \right]} \qquad (19a)$$

where $k_B T_C = -2\beta$ is the energy gain per Cooper pair upon joining the condensate. States for which $0 < \Delta E_i < -2\beta$ (see eq. 15a) are said to be coherent, and may contribute to the condensate in the form of virtual vortex-anti-vortex pairs (*vide infra*). $\Delta E_i = -2\beta$ and $\Delta E_i = -4\beta$ reflect disconnected Cooper pairs and dissociated super-atoms, respectively. In particular, if all excited coherent states on the ring displays energies greater than the Cooper pair evaporation energy -2$\beta$,i. e $E_k - E_0 > -2\beta$, this implies

$$M = \left[\frac{N}{2}\right] \times \frac{\frac{1}{e^{\frac{2\beta}{k_B T}} - 1}}{\frac{1}{e^{\frac{2\beta}{k_B T_C}} - 1}} = \left[\frac{N}{2}\right] \times \frac{e - 1}{e^{\frac{2\beta}{k_B T}} - 1} \qquad (19b)$$

Thus, the temperature dependence of the super-fluid pair number-density becomes

$$N_s^0 = \frac{\left[\frac{N}{2}\right] - M}{\left[\frac{N}{2}\right]} = 1 - \frac{\sum_i \left[\frac{1}{e^{\frac{\Delta E_i}{k_B T}} - 1}\right]}{\sum_i \left[\frac{1}{e^{\frac{\Delta E_i}{k_B T_C}} - 1}\right]} \qquad (20)$$

and the expression for the ground state superfluid density thus becomes

$$n_s^0(\vec{r}) = \left[\frac{N}{2}\right] \times N_s^0 \times \left| \frac{1}{\sqrt{N - 2M}} \sum_{\substack{i=1 \\ i \in BEC}}^{N-2M} \phi_i(\vec{r}) \right|^2 \qquad (21)$$

where it is implied that the evaporations of Cooper pairs result in gradually shrinking rings of condensed Cooper pairs. In reality, $n_s^0(\vec{r})$ reflects one out of likely several nearest-neighbor arrangements of super-atoms hosting delocalized Cooper pairs, coexisting with disconnected individual Cooper pairs. In such a scenario, these BEC islands could be understood to achieve extended phase coherence in two dimensions by Josephson coupling.

In order to demonstrate the usefulness of the above ansatz we make contact with vortex-antivortex pairs scenarios of HTSC. Let the eigenstates Eq. (15b) represent ground state and collective excitations of the BEC. Then we may approximate the BEC wave function by an axially symmetric eigenstate for particles-on-a-ring, i.e.

$$\Psi_{\pm|\mathrm{m}|}(\vec{r}) = \sqrt{n_s^{\pm|m|}(\vec{r})} \times \left( \frac{1}{\sqrt{N-2M}} \sum_{\substack{j=1 \\ j \in BEC}}^{N-2M} \exp[i \cdot (\pm)|m| \cdot \pi \tfrac{j-1}{N-2M}] \phi_j \right)$$

$$\propto \sqrt{n_s^{\pm|m_l|}(\vec{r})} \times \frac{1}{\sqrt{2\pi}} e^{\pm i|m_l|\varphi} \qquad (22a)$$

where the doubly-degenerate states Eqs (15a, 15b) are then taken to be the real and imaginary components of two complex wave functions complementary to a vortex anti-vortex pair. These coherent excited states, corresponding to the angular moments $\pm|m_l|\hbar$, comprise non-dissipative contributions to the condensate *if and only if* accessed as *virtual* vortex-antivortex pairs. When originating e.g. from thermal excitations, these collective excitations cause phase slips which translate into condensate dissipation. The angular moments of the coherent excited states are proportional to the corresponding orbital magnetic moments of the superfluid excitation with macroscopic magnetizations

$$\mathcal{M}_{\pm}^{m_l} = \mp n_s^0(\vec{r}) \times |m_l|\hbar \frac{2e}{2m_e} \qquad (22b)$$

Finally, for completeness, we make explicit contact with the London equations by identifying the above condensate wave function (compare Eq. 22a) with its general wave-particle dual form

$$\Psi_0(\vec{r}) = \sqrt{n_s(\vec{r})} \times e^{\frac{iS}{\hbar}} \qquad (23a)$$

where $S$ is the classical action for a single Cooper pair belonging to the condensate. Thus for example the chemical potential -2$\beta$ for adding a Cooper pair to the condensate comes out as

$$\mu = -\frac{\partial S}{\partial t} \qquad (23b)$$

Moreover, from the familiar expression for the current density in an external magnetic field

$$\vec{j} = \frac{ie\hbar}{2m_e}\left(\Psi_0^*\vec{\nabla}\,\Psi_0 - \Psi_0\vec{\nabla}\Psi_0^*\right) - \frac{2e^2}{m_e}\vec{A}\cdot\Psi_0^*\Psi_0 \qquad (23c)$$

we obtain

$$\vec{j} = -\frac{2e^2}{m_e}\Psi_0^*\Psi_0\left(\frac{\vec{\nabla}S}{2e} + \vec{A}\right) = -\frac{n_s^0(\vec{r})e^2}{m_e}\left(\frac{\vec{\nabla}S}{2e} + \vec{A}\right) \qquad (23d)$$

where $\vec{\nabla}S$ is the local canonical momentum of a single Cooper pair belonging to the condensate. The London equations, finally, may be derived from Eqs. (23b,23d) and Maxwell's equations. In particular, Eq. (22b) is a special case of Eq. (23d). This completes the discussion on the resulting HTSC ground state projected on the manifold of electronic states of the superatoms subsystem.

## 4. Complementarities of super-atoms and local AFM perspectives

The complementary perspective to that of the super-atom excitations may also be taken, i.e. approaching the superconductivity from the AFM perspective. Hence, we say that virtual magnons mediating the virtual BEC among super-atoms by propagating the shared pair-states, necessarily requires virtual BEC among magnons if and only if the phase difference between the corresponding two order parameters is locked compare Eq. 3. From a super-atoms excitations perspective we may write the chemical potential $\mu$ for real-space Cooper pairs joining the BEC in terms of probability densities for local pair- and pair-broken states

$$\mu_{BEC}^{CCP} = -2\beta = 2\cdot u^2 v^2 J_{local} \qquad (24)$$

where

$$n_P = v^2 \ , \ n_{PB} = u^2 \qquad (25)$$

such that

$$\mu_{BEC}^{CCP} = 2n_P n_{PB} J_{local} = 2n_P(1-n_P)J_{local} = 2n_{PB}(1-n_{PB})J_{local} \qquad (26)$$

Now, in order to articulate the complementary AFM perspective, let *m* be the virtual magnon probability density caused by gain in AFM "electron correlation entropy", and $\delta_{local}$ a local cluster pair-breaking excitation energy, See Fig. 2A. Analogous to Eqs. 26, we write:

$$\mu_{BEC}^{magnon} = 2 \cdot m(1-m)\delta_{local} \qquad (27)$$

Because BEC in the super-atoms channel is understood to require magnons BEC, we equate the two expressions, Eq. 26 and Eq. 27, i.e. $\mu_{BEC}^{CCP} = \mu_{BEC}^{magnon}$ to obtain

$$\frac{J_{local}}{\delta_{local}} = \frac{m}{n_{PB}} \cdot \frac{1-m}{n_P} \qquad (28)$$

Similarly, let $\delta_{local} = J_{local}$ be the resonance requirement for the applicability of Eq. 3, which implies that for every $\delta_{local}$ there is a $J_{local}$ up to $J_{\max}$ where the latter corresponds to the superexchange interaction in the undoped system. This is equivalent to stating that the "electron correlation entropy" in the AFM channel is equivalent to that in the super-atom channel. The distributions of $\delta_{local}$ in the cuprates would result from the disorder in the A-site rendering $J_{local}$ inhomogeneous due to bond lengths changes in adjacent in-plane Cu-O bonds [23]. The upper bound to the chemical potential for condensation of Cooper pairs is arrived at (cf. Eq. 24, for $u = v = \frac{1}{\sqrt{2}}$)

$$\mu_{BEC} = \tfrac{1}{2}\delta_{local} = \tfrac{1}{2}J_{local} \qquad (29)$$

Let $J_{local} = J_{\max}$ and assume $J_{\max}$ to reflect the Neél temperature in the same way as $\mu$ reflects $T_C$, we obtain

$$T_C^{\max} \sim \frac{1}{2} T_N \tag{30}$$

Hence, the equivalence of two complementary descriptions of the HTSC has been underlined by equating the super-atoms perspective to that of the antiferromagnet. Indeed, many of the phenomenological properties of the cuprate superconductors may be traced back to the bands which accommodate the AFM. Here, the two complementary components come together in the HTSC phenomenon, comprising the non-adiabatic coupling of AFM and superatom manifolds of electronic states.

The close relationship between the Neél temperature of the undoped system, and the maximum critical temperature for superconductivity is repeatedly emphasized. This interpretation is in line with recent reports of possible local superconducting correlations above $T_C$ in the cuprates [40]. It is gratifying to note here that room-temperature Bose-Einstein condensation of magnons, obtained upon pumping, has recently been reported [41].

## 5. Concluding remarks

The aim of the present study is to offer a compact possible understanding of high-$T_C$ superconductivity, which appeals equally to both physical and chemical intuition. The above Hückel-type RVB-BEC exercise is complementary to our previous real-space BCS formulations [7,8,31]. Thus two perspectives on one and the same quantum chemical understanding of HTSC is offered. The essential elements comprise two *a priori* disjoint electronic sub-systems, each exhibiting a manifold of electronic states in the vicinity of the common Fermi energy. Resonant coupling between the two sub-systems enforces the effective non-local phase rigidity.

A rationale for the robustness of the HTSC ground state is proposed based on local maximum "electron correlation entropy" at the expense of non-local phase rigidity. Possible evidence for such resulting synergy to emerge from *a priori* separated subsystems is provided in a recent ARPES study of the pseudo-gap state [42], which reports particle-hole asymmetry as well as translational symmetry breaking. The connection to the understanding presented here is made in [7,31,37], i.e. in terms of inter-band charge transfer in conjunction with formation of super-atoms. Consequently it is implied that the pseudo-gapped state observed e.g. by Hashimoto et al. [42] is the precursor to the superconducting state and not some competing phase. Indeed, the results of the present study is in line with a recent ARPES study [43], which reports a temperature, "$T_{pair}$", extracted from the onset of non-linear loss of DOS at the anti-nodes of the Bi-cuprates. Thus, "$T_{pair}$" was taken to signify formation of Cooper pairs above the critical temperature for superconductivity.

***Computational details***

*The CASTEP [44] program package within the Material Studios framework [45] was utilized and the GGA PBE functional [46] employed. Core electrons were described by ultra-soft pseudo-potentials, in conjunction with 300 eV cut-off energy. Gamma point calculations are presented through-out employing* $4a_0 \times 4a_0$ *super cells.*

**References**

1. J. G. Bednorz, K. A. Müller, *Z. Phys.* **B64** (1986) 189.

2. S.-C. Zhang, *Science,* **275 (**1997) 1089.

3. C. M. Varma, Z. Nussinov and W. van Saarloos, *Phys. Rep.* **361** (2002) 267.

4. S. Sachdev, Quantum Phase Transitions (Cambridge Univ. Press, Cambridge, 1999).

5. see D.J. Scalapino, Handbook of High Temperature Superconductivity, Chapter 13, Eds. J.R. Schrieffer and J.S. Brooks (Springer,New York, 2007).

6. S. Brehm, E. Arrigoni, M. Aichhorn and W. Hanke, *Europhysics Letters* **89**, 27005 (2010).

7. I. Panas, *J. Phys. Chem.* **B103** (1999) 10767.

8. I. Panas, , American Physics Society Proceedings 483 (1999) 85.

9. I. Panas, A. Snis, and F. Bawa, *J. Low Temp. Physics*, **117** (1999) 419.

10. I. Panas, and R. Gatt, *Chem. Phys. Letters*, **259** (1996) 241.

11. I. Panas, and R. Gatt, *Chem. Phys. Letters*, **259** (1996) 247.

12. I. Panas, and R. Gatt, *Chem. Phys. Letters*, **266** (1997) 410.

13. I. Panas, and R. Gatt, *Chem. Phys. Letters*, **270** (1997) 178.

14. K. McElroy et al., *Phys. Rev. Lett*. **94** (2005) 197005.

15. J. E. Hoffman *et al*. "A four unit cell periodic pattern of quasi-particle states surrounding vortex cores in $Bi_2Sr_2CaCu_2O_{8+\delta}$." *Science* **295**, 466–469;

16. W. D. Wise *et al*. "Charge-density-wave origin of cuprate checkerboard visualized by scanning tunnelling microscopy". *Nature Phys.* **4**, 696–699 (2008)

17. W. D. Wise, *et al*. Imaging nanoscale Fermi-surface variations in an inhomogeneous superconductor. *Nature Phys.* **5**, 213–216 (2009).

18. P. Bourges, in *Gap Symmetry and Fluctuations in High Temperature Superconductors* ed. by J. Bok, G. Deutscher, D. Pavuna and S.A. Wolf (Plenum, 1998).

19. S.W. Cheong, G. Aeppli, T.E. Mason, H. Mook, S.M. Hayden, P.C. Canfield, Z. Fisk, K.N. Clausen, J.L. Martinez, *Phys. Rev. Lett.* **67** (1991) 1791.

20. H. F. Fong, B. Keimer, D. L. Milius, I. Aksay, A. *Phys. Rev.Lett.* **78** (1997) 713.

21. H. A. Mook, Pengcheng Dai, S. M Hayden, G. Aeppli; T. G Perring, F. Dogan, *Nature* **395** (1998) 580.

22. R. Ofer, G. Bazalitsky, A. Kanigel, A. Keren, A. Auerbach, J. S. Lord, and A. Amato, *Phys. Rev.* **B 74** (2006) 220508(R).

23. W. B. Gao, Q. Q. Liu, L. X. Yang, Y. Yu, F. Y. Li, C. Q. Jin, and S. Uchida *Phys. Rev.* **B 80**, 94523 (2009)

24. C. C. Tsuei, J. R. Kirtley, C. C. Chi, Lock-See Yu-Jahnes, A. Gupta, T. Shaw, J. Z. Sun, and M. B. Ketchen, *Phys. Rev. Lett.* **73** (1994) 593.

25. G. Levy, M. Kugler, A. A. Manuel, and Ø. Fischer, Phys. Rev. Lett. **95** (2005) 257005.

26. T. Timusk, and B. Statt, "The pseudogap in high-temperature superconductors: An experimental survey". *Rep. Prog. Phys.* **62** (1999) 61–122.

27. Norman, M. R. *et al.* "Destruction of the Fermi surface in underdoped high-Tc superconductors." *Nature* **392**, (1998) 157–160.

28. J. Meng, *et al.* "Coexistence of Fermi arcs and Fermi pockets in a

high-*T*c copper oxide superconductor", *Nature* **462**, (2009) 335-338.

29. S. Margadonna, Y. Takabayashi, Y. Ohishi, Y. Mizuguchi, Y. Takano, T. Kagayama, T. Nakagawa, M. Takata, and K. Prassides, *Phys. Rev. **B*** **80** (2009) 064506.

30. Kanihara, Y., Watanabe, T., Hirano, M., Hosono, H., et al.: *J. Am. Chem. Soc.* **130**, (2008) 3296–3297; Z.Wei, H. Li , W.L. Hong, Z. Lv, H.Wu, X. Guo, K. Ruan, *J Supercond Nov Magn* **21** (2008) 213.

31. I. Panas, *Phys. Rev.* **B82** (2010) 064508.

32. J. Bardeen, L. N. Cooper, and J. R. Schrieffer, *Phys. Rev.* **106** (1957)162.

33. P.W. Anderson, *Mater. Res. Bull*. **8** (1973) 153.

34. P.W. Anderson, *Science*, **235** (1987) 1196.

35. P.W. Anderson, Baskaran, Z. Zhou, and T. Hsu, *Science* **58** (1987) 2790.

36. F. Steglich, J. Aarts, C.D. Bredl, W. Lieke, D. Meschede, W. Franz, and H. Schäfer, *Phys. Rev. Lett*. **43** (1979)1892-1896.

37. I. Panas, *Phys. Rev.* **B 83** (2011) 024508.

38. C.E. Shannon, *Bell System Technical Journal*, **27** (1948) 379–423, 623-656.

39. A. Mesaros K. Fujita, H. Eisaki, S. Uchida, J. C. Davis, S. Sachdev, J. Zaanen, M. J. Lawler, Eun-Ah Kim, *Science* **333**, (2011) 426.

40. Lu Li, Yayu Wang, Seiki Komiya, Shimpei Ono, Yoichi Ando, G. D. Gu, and N. P. Ong, *Phys. Rev.* **B 81**, 054510 (2010).

41. S. O. Demokritov, V. E. Demidov, O. Dzyapko, G. A. Melkov, A. A. Serga, B. Hillebrands, and A. N. Slavin, *Nature*, **443** (2006) 430.

42. M. Hashimoto et al. *Nature Physics*, **6**, (2010) 414 - 418

43. T. Kondo *et al.* "Disentangling Cooper-pairs formation above the transition temperature from the pseudogap state in the cuprates", *Nature Physics* 7, (2011)

44. S. J. Clark, M. D. Segall, C. J. Pickard,P. J. Hasnip, M. J. Probert, K. Refson, M. C. Payne, *Zeitschrift Für Kristallographie*, **220** (5-6) (2005) 567.

45. Materials Studio 5.0, Accelrys Inc., Simulation software.

46. J. P. Perdew, K. Burke, and M. Ernzerhof, *Phys. Re*v. *Lett.* **77** (1996) 3865.